\documentclass[aip,rsi,reprint,floatfix]{revtex4-1}

\usepackage{graphicx}
\usepackage{epstopdf}
\usepackage{color}
\usepackage{textcomp} 
\usepackage{amssymb}
\usepackage[T1]{fontenc}
\newcommand{\AgSi}{Ag/Si$(111)-(\sqrt3\times\sqrt3)\rm R30^{\circ}$}
\newcommand{\Si}{Si$(111)-(7\times7)$}
\begin{document}

\title{Scanning tunneling potentiometry implemented into a multi-tip setup by software}
\author{F. L\"upke}
\affiliation{Peter Gr\"unberg Institut (PGI-3), Forschungszentrum J\"ulich, D-52425 J\"ulich, Germany}
\author{S. Korte}
\affiliation{Peter Gr\"unberg Institut (PGI-3), Forschungszentrum J\"ulich, D-52425 J\"ulich, Germany}
\author{V. Cherepanov}
\affiliation{Peter Gr\"unberg Institut (PGI-3), Forschungszentrum J\"ulich, D-52425 J\"ulich, Germany}
\author{B. Voigtl\"ander}
\altaffiliation[Electronic mail: ]{\urlstyle{same}\url{b.voigtlaender@fz-juelich.de}}
\affiliation{Peter Gr\"unberg Institut (PGI-3), Forschungszentrum J\"ulich, D-52425 J\"ulich, Germany}
\date{\today}


\begin{abstract}
{We present a multi-tip scanning tunneling potentiometry technique that can be implemented into existing multi-tip scanning tunneling microscopes without installation of additional hardware}. The resulting setup allows flexible {\it in situ} contacting of samples under UHV conditions and subsequent measurement of the sample topography and local electric potential with resolution down to \AA\ and \textmu V, respectively. The performance of the potentiometry feedback is demonstrated by thermovoltage measurements on the \AgSi\ surface by resolving a standing wave pattern. Subsequently, the ability to map the local transport field as a result of a lateral current through the sample surface is shown on \AgSi\ and \Si\ surfaces.
\end{abstract}
\maketitle

\section{Introduction}
In recent years, multi-tip scanning tunneling microscopes (STM) have become popular instruments \cite{Hobara2007, Kim2007, Nakayama2012, Bannani2008}. The most common experiments reported in literature using those instruments are multi-probe measurements that allow to determine the microscopic electrical properties of the samples under investigation. However, such devices can be further extended with scanning tunneling potentiometry (STP) techniques which allow to map the electric potential on the nanoscale with spatial and potential resolution of few \AA\ and \textmu V, respectively \cite{Pelz1989, Druga2010, Bannani2008}. Combination of STP with multi-probe point measurements creates a powerful tool for the electrical sample characterization capable to give insight into fundamental transport properties, such as the influence of defects on the local electric transport. This applies especially for surface dominated transport as present e.g. in topological insulators and allows the investigation of transport phenomena such as the Landauer dipole \cite{Feenstra1998,Willke2015}.
\begin{figure}[!b]
\includegraphics{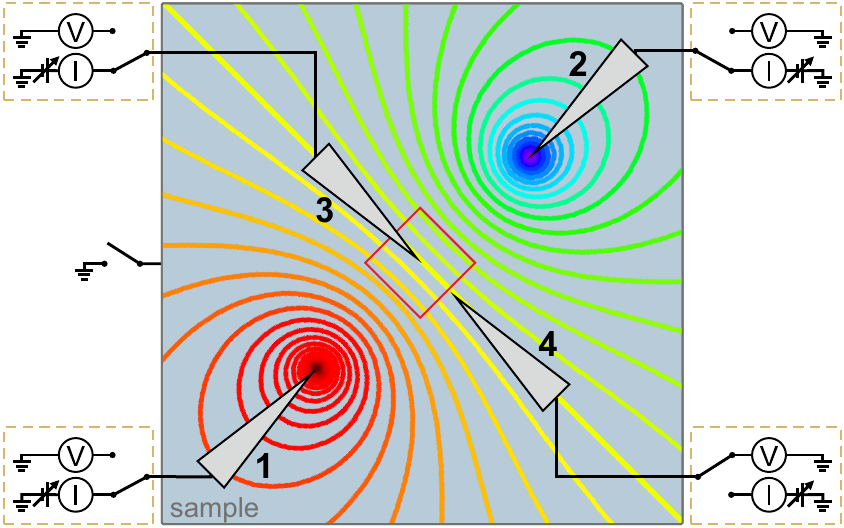}
\caption{(Color online) Schematic top view of the four-tip STP setup. Tips 1 and 2 are in contact to the sample surface and inject a lateral current represented by the colored equipotential lines. Tip 3 is in tunneling contact and is scanned across the surface. The scan area is indicated as red square (largely exaggerated). Tip 4 is in contact to the sample surface close to the scanning area and serves as reference voltage probe.}
\label{1}
\end{figure}
Since the invention of STP by Murlat and Pohl \cite{Muralt1986} several different technical implementations of STP have been reported mostly in single-tip STM setups \cite{Druga2010, Rozler2008} for which the transport field is usually applied to macroscopic sample contacts up to several mm away from each other. As a result, the current density in the region of measurement can be low, resulting in a bad signal-to-noise ratio of the measured electric potential distribution throughout the sample surface. To circumvent this problem, patterning metallic contacts on the sample surface was reported to increase the current density across the structure under investigation  \cite{Wang2013,Willke2015}. However,  the patterning process often results in contamination of the sample surface.

In contrast, multi-tip STM setups have the advantage of flexible {\it in situ} contacting the sample with several electrodes, e.g. molecular beam epitaxy grown films and nanostructures without additional sample processing \cite{Bannani2008,Hobara2007,Kim2007}. As a result, by variation of the spacing of the electrodes the current density in the surface and also through the bulk can be measured and controlled \cite{Just2015}. Furthermore, the direction of injected current can be varied by repositioning of the electrodes enabling the quantification of transport anisotropies \cite{Kanagawa2003}. 

Multi-tip STP implementations have been reported that facilitate the above mentioned advantages over single-tip setups \cite{Bannani2008}. {However, these implementations are based on additional hardware such as external power sources and feedback electronics that need to be added to existing setups in order to enable potentiometry measurements \cite{Druga2010, Bannani2008}. In contrast, we present here a software based multi-tip STP technique that can be readily implemented into existing multi-tip setups without installation of additional hardware}.

\section{Experimental Setup}

{We use a home-built room temperature four-tip STM in Besocke/Beetle-based design with electrochemically etched tungsten tips}. Each tip has an individual ring for xy-motion and a tube piezo for z-motion \cite{Cherepanov2012, mProbes2015}. Atomic resolution on \Si\ is achieved with each tip. The STM is equipped with a scanning electron microscope (SEM) for precise positioning of the individual tips.

Each of the four tips is connected to one of four instances of equivalent electronics which can be switched between current or voltage measurement, as also done in other four-tip STM setups \cite{Hobara2007}. For current measurement, the voltage applied at each tip can be set individually. Furthermore, the sample can be either on floating or ground potential. In STP measurements three of the four tips are in current probe mode, two current injection tips (tip 1 and 2) and the scanning tip (tip 3). The fourth probe is operated in voltage probe mode and is used as a reference voltage probe (tip 4). This tip is optional such that the presented implementation is also applicable for three-tip STM setups. 
A schematic of the setup is shown in FIG.~\ref{1}.

The four biasing electronics units are connected to a commercial multi-tip STM controller with individual digital-to-analog converters (DAC) outputs to address the bias voltage for each tip and analog-to-digital converters (ADC) to read out the current or voltage at each tip.

\section{Biasing electronics}
\begin{figure}[!ht]
 \includegraphics{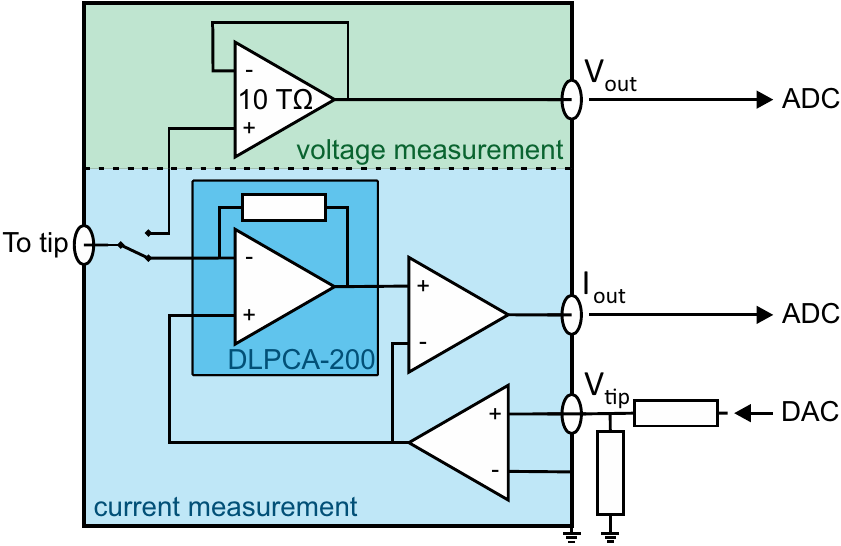}
 \caption{(Color online) Schematic of the biasing electronics for each tip. In the upper part the voltage measurement is performed by a voltage follower circuit. In the lower part the current measurement schematic is shown. The bias voltage $V_{\rm tip}$ is applied via the differential input in front of which the optional voltage divider for the scanning tip is located.}
 \label{2}
 \end{figure}

Figure~\ref{2} shows a schematic of the biasing electronics which is typical for a multi-tip setup \cite{Hobara2007}. The present setup is based on a FEMTO DLPCA-200 current amplifier \cite{Femto2015} and allows variable-gain current measurement or voltage measurement. For current measurement the tip is connected to the input of the current amplifier which is floating on bias potential $V_{\rm tip}$. From the output of the current amplifier, the bias voltage is subtracted before the signal is fed to the output $I_{\rm out}$. For voltage measurement the tip is connected to a high input impedance ($10\rm\;T\Omega$) voltage follower with its output connected to the $V_{\rm out}$ output of the electronics.

In high resolution STP measurements, the control of $V_{\rm tip}$ of the scanning tip in the range of \textmu V is required. This can be achieved by a voltage divider (e.g. $1:100$) which is applied just in front of the $V_{\rm tip}$ input of the scanning tip biasing electronics. This decreases the dynamic range of the voltage which can be applied to this tip accordingly. Not to be limited by this, the topography tip-sample tunneling bias voltage is applied to the sample rather than the tip as described below.
  
\section{Software Implementation}
 \begin{figure}[!h]
 \includegraphics{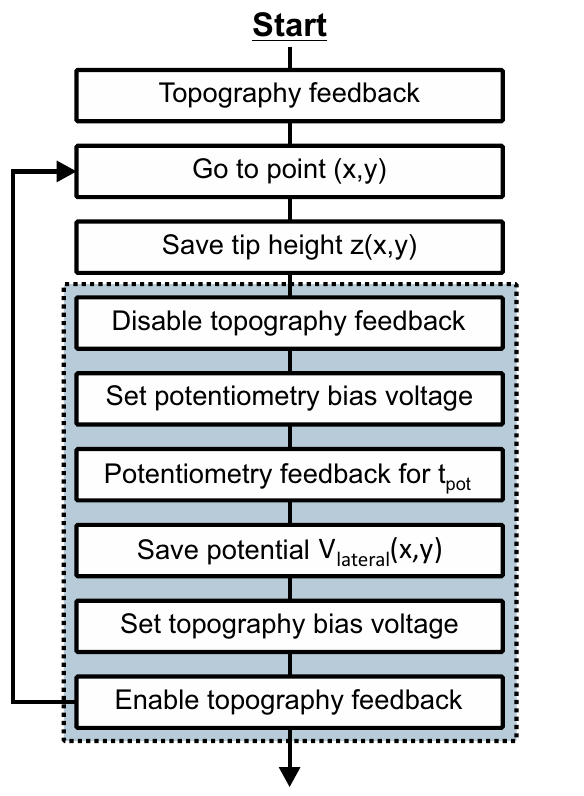}  
 \caption{(Color online) Potentiometry measurement flow chart. Highlighted by the dotted rectangle is the additional procedure needed to acquire the sample potential in comparison to conventional STM.}
 \label{3}
 \end{figure}
  \begin{figure}[!ht]
  \includegraphics{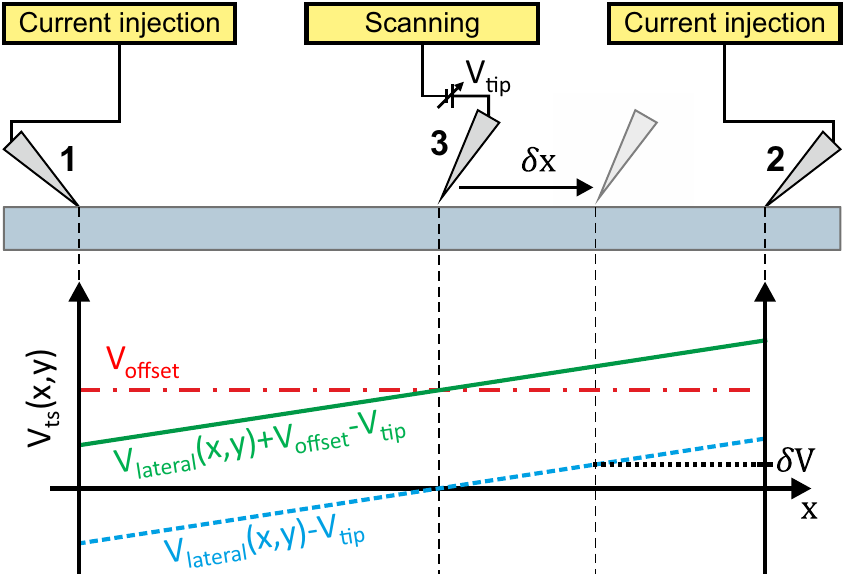}  
  \caption{(Color online) Schematic of the potentiometry measurement. Tips 1 and 2 are in contact to the sample injecting a current. Tip 3 is in tunneling contact and is scanned across the sample surface. The dashed vertical lines indicate the position of the tips, respectively. In the line graph the tip-sample voltage $V_{\rm ts}$ is plotted as a function of the lateral coordinate $x$. During topography feedback, the solid green line holds. For potentiometry feedback $V_{\rm ts}$ is shifted by $V_{\rm offset}$ resulting in the dashed blue line. Moving the scanning tip by an amount $\delta x$ results in a change of the potential under the tip by $\delta V$. Note: The movement of the scanning tip is largely exaggerated in this graph.}
  \label{4}
  \end{figure}

The multi-tip STP implementation presented here is based on the interrupted feedback loop single-tip STP implementation reported in ref. \onlinecite{Druga2010}. In this technique the surface is scanned point-by-point as in regular STM measurements. However, at each scanning point after the topography value $z(x,y)$ is acquired, the tip is held at constant height and the bias voltage between scanning tip and sample is shifted. Now the potentiometry feedback is run for a time $t_{\rm pot}\approx1-50\rm\;ms$. The final value of the potentiometry feedback is saved as the sample potential at the momentary position of the scan $V_{\rm lateral}(x,y)$. After restoring the topography bias voltage, the topography feedback is re-enabled and the scan is continued by moving to the next scan position \cite{footnote1}. The flow chart of the measurement procedure is shown in FIG.~\ref{3}. In this way topography and potentiometry signal are acquired point-by-point across the sample surface. Depending on the feedback time $t_{\rm pot}$ and the number of scan points a whole scan takes up to several hours.

The potentiometry feedback is a PI-feedback algorithm just like the one used for the topography feedback. However, it adjusts the tip voltage as control variable instead of the tip height (in topography feedback) and the process variable set point (i.e. the tunneling current) is set to $I_{\rm t}=0\rm\;nA$ during the potentiometry feedback. The corresponding feedback constants determine the bandwidth of the potentiometry feedback and have to be adjusted for each measurement.

\section{Measurement procedure} 
In order to perform a potentiometry measurement, first the sample (connected to ground) needs to be contacted by the two current injecting tips (tip 1 and 2). After coarse approaching all four tips to the sample surface, the current injecting tips are brought into contact to the surface under tunneling conditions of e.g. $V_{\rm tip}=1\rm\;V$; $I_{\rm t}=10\rm\;pA$. One after the other, the current injection tips are advanced further by a few nm with disabled topography feedback until a stable current ($\lesssim$1\% variation per minute) of up to several \textmu A is measured between that tip and the sample. Typically, there is a sharp increase of the current as the tips come into contact to the surface after which the tips should not be advanced further to prevent damage to the tips or the sample surface. However, if at this point the resulting current is unstable, advancing the tips a bit further can help to stabilize the contact.

Next, the optional reference voltage probe (tip 4) is brought into contact to the sample surface in the same way (in current measurement mode). When a stable contact is achieved, the electronics of this tip is switched to voltage measurement. This tip serves to measure the reference voltage which can be used during the measurement preparations as described below but also to compensate small fluctuations in the current injecting contacts during measurements \cite{Bannani2008}: {Changes of the contact resistance in the current injecting contacts result in a change of the sample potential at the point of current injection. This will directly influence the measured potential at the position of the scanning tip and can be compensated by subtracting the potential of a static reference voltage probe from the measured potential of the scanning tip.}

Now the sample backside contact is disconnected such that the electric potential of the sample depends only on the potential of the two current injecting tips.

To approach the scanning tip (tip 3), the voltage of both injection tips is set to 0 V while that of the scanning tip is set to e.g. $V_{\rm tip}=1\rm\;V$. Then the tip is approached to the sample surface until it is in tunneling contact ($I_{\rm t}\sim 10\rm\;pA$). Subsequently, a topography scan should be performed to check if the scan area is suitable for potentiometry measurements.

To apply a transport field, first the scanning tip (tip 3) is retracted and its voltage is set to $V_{\rm tip}=0\rm\;V$. Now the voltage of the injection tips is set such that a lateral current $I_{\rm t}\gtrsim1$ \textmu A results in a potential distribution $V_{\rm lateral}(x,y)$ over the sample surface. Subsequently, $V_{\rm lateral}(x,y)$ is shifted until the sample potential under the scanning tip is approximately $0\rm\;V$. This can be done using the information of the voltage probe. In order to adapt $V_{\rm lateral}(x,y)$ at the position below the scanning tip, the voltage of both current injecting tips is shifted identically, maintaining the injected current \cite{footnote2}. 
Next, the bias voltage for the topography measurement $V_{\rm offset}$ is added to the voltage of the current injecting tips. 
For high resolution measurements the voltage divider is now applied to $V_{\rm tip}$ of the scanning tip. Following the described preparations, the tunneling contact of tip 3 is re-engaged and the STP scan can be started. Figure~\ref{4} shows a schematic of the potentiometry measurement setup with the voltages applied to the tip and sample indicated, respectively.

In STP, the voltage difference between the scanning tip and sample $V_{\rm ts}(x,y)$ usually referred to as bias voltage at the scan position $(x,y)$ is the superposition of the lateral transport field, the offset voltage and the momentary voltage applied to the tip during topography feedback $V_{\rm ts}(x,y)=V_{\rm lateral}(x,y)+V_{\rm offset}-V_{\rm tip}$ (solid green line in FIG.~\ref{4}). When starting the STP scan, after acquiring the topography at the momentary scan position, the topography feedback is deactivated and the injection tip voltages are shifted by $-V_{\rm offset}$ leaving only $V_{\rm lateral}(x,y)$ applied to the sample resulting in $V_{\rm ts}(x,y)=V_{\rm lateral} (x,y)-V_{\rm tip}$ (dashed blue line in FIG.~\ref{4}). Subsequently, the potentiometry feedback loop is activated and adjusts $V_{\rm tip}$ to the local sample potential under the tip $V_{\rm tip}=V_{\rm lateral}(x,y)$. After the feedback time $t_{\rm pot}$, the potentiometry feedback is deactivated. $V_{\rm tip}$ is recorded as the sample potential at this scan position and is maintained until the next activation of the potentiometry feedback loop. After restoring the topography bias voltage $V_{\rm ts}(x,y)=V_{\rm lateral}(x,y)+V_{\rm offset}-V_{\rm tip}$, the topography feedback is re-activated and the tip is moved to the next scan position. 
 
Note that, for neighboring scan positions ($x\rightarrow x+\delta x$), the local sample potential varies by an amount of $\delta V$ during the topography feedback. In detail, during the topography feedback it holds $V_{\rm ts} (x+\delta x,y)=V_{\rm offset}+\delta V$. However, in usual measurements $\delta V\ll V_{\rm offset}$ such that the change in $V_{\rm ts}$ for neighboring scan positions can be neglected \cite{Druga2010}.

For the potentiometry feedback to work in an optimal way it is recommended that the tunneling $I-V$ curve of the scanning tip is linear, such that a change in the control variable (tunneling voltage) will have a linear effect on the process variable (tunneling current) \cite{Hamada2012}. Furthermore, due to the fundamental noise limit of the potential measurement which is the sum of thermal (Johnson) noise and the amplifier noise the limit of the potentiometry resolution is \cite{Rozler2008} 
\begin{equation}
V_{\rm Limit}=\sqrt{V_{\rm Johnson}^2+V_{\rm Detector}^2} 
\end{equation}
with
\begin{equation}
V_{\rm Johnson}=\sqrt{4k_B T \Delta f V_{\rm ts}/I_{\rm t}}
\end{equation}
and
\begin{equation}
V_{\rm Detector}=N_{\rm Detector}\sqrt{\Delta f}V_{\rm ts}/I_{\rm t}.
\end{equation}
Here $N_{\rm Detector}=4\rm\;fA/\sqrt{Hz}$ is the noise level of the current amplifier \cite{Femto2015}.
Low tunneling voltages and high tunneling current setpoint depending on the bandwidth $\Delta f$ of the setup are necessary for low noise potentiometry measurement \cite{Rozler2008}. For transport measurements with several mV of voltage drop across the scan area, we determine $\Delta f<50\rm\;Hz$ as the feedback loop speed for which a good signal-to-noise ratio is achieved while a relatively high scanning speed is maintained. For this bandwidth, the noise limit for three different tunneling resistances is given in TABLE~\ref{Table 1}.

\begin{table}[h!]
\begin{tabular}{r|r}
$V_{\rm ts}/I_{\rm t}$  & $V_{\rm Limit}$ \\ \hline
 $10\rm\;G\Omega$  & $<300\rm\;\text{\textmu} V$ \\ 
\hline $1\rm\;G\Omega$ & $<40\rm\;\text{\textmu} V$ \\ 
\hline $100\rm\;M\Omega$ & $<10\rm\;\text{\textmu} V$
\end{tabular}
\caption{Noise limit for transport measurements in the present setup at three different tunneling junction resistances and a bandwidth of $\Delta f<50\rm\;Hz$. \label{Table 1}}
\end{table}

\section{Results}
\subsection{Thermovoltage on \AgSi} 
 \begin{figure}[!h] 
 \includegraphics{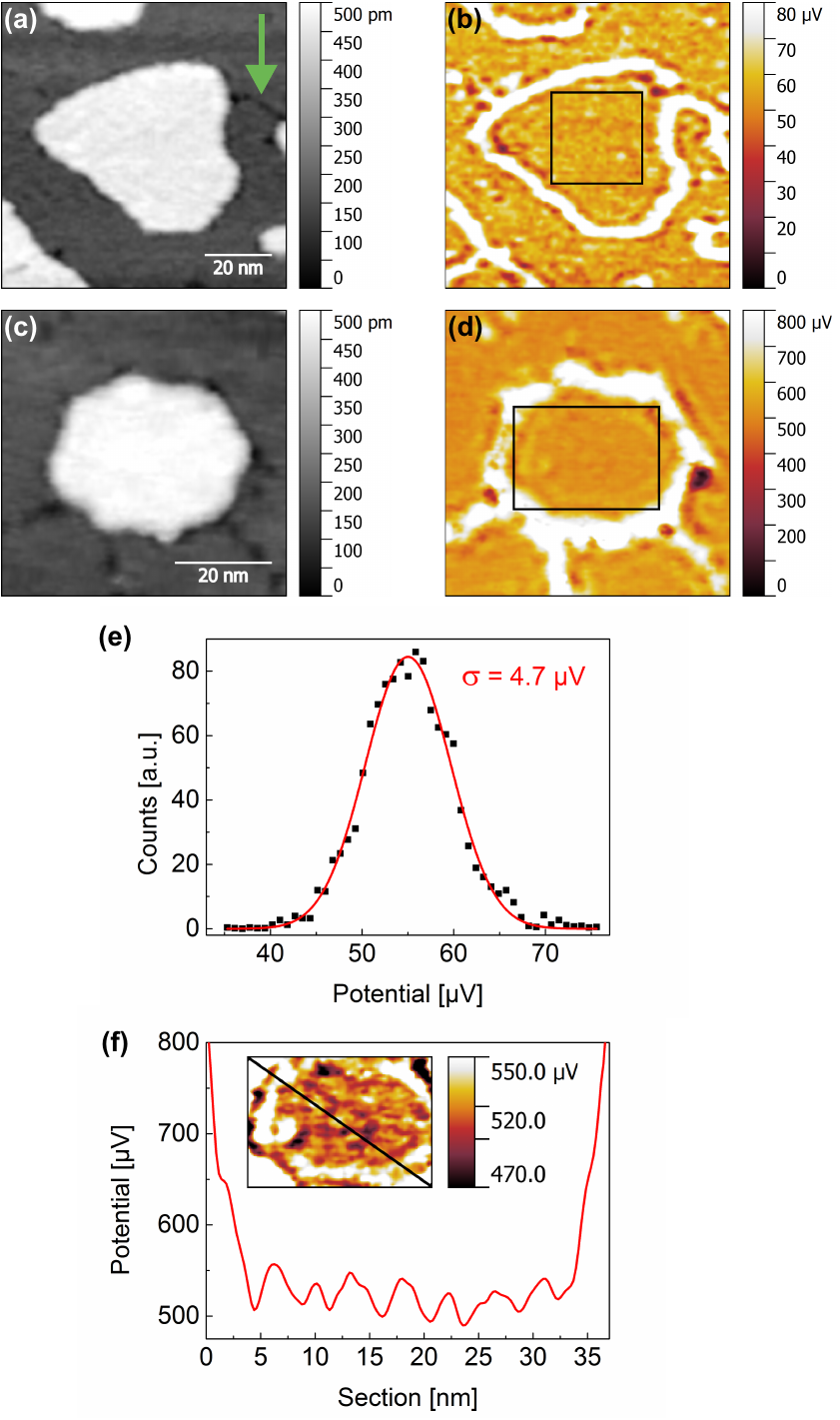}
 \caption{(Color online) Thermovoltage measurements on \AgSi. (a) Topography of an island on the \AgSi\ surface at $V_{\rm offset}=10\rm\;mV$ and $I_{\rm t}=0.1\rm\;nA$ with a domain boundary indicated by the green arrow. (b) The corresponding potential map shows an enhanced potentiometry signal at the edges of the islands and domain boundaries. The potentiometry feedback time is $20\rm\;ms$ at a bandwidth of $\Delta f=2 \rm\;Hz$. (c) Topography of an island with the scanning tip under laser irradiation. {(d) Corresponding potential map of (c). The amplitude of the potentiometry signal is increased by a factor of around 10 compared to (b). (e) Potential distribution of a featureless surface area (black square in (b)). The standard deviation is $\sigma=4.7\rm\;$\textmu V. (f) Section along the black line in the inset that corresponds to the black rectangle in (d). A standing wave pattern with a periodicity of $4.29(14)\rm\;nm$ is observed. The section is averaged over 3 lines.}}
 \label{5}
 \end{figure}

The \AgSi\ reconstruction is a well-studied sample system when it comes to scanning tunneling microscopy, spectroscopy and also potentiometry \cite{Druga2010, Bannani2008}. Therefore, this system can be used as a benchmark system to test the performance of the presented STP implementation. For the preparation of the \AgSi\ surface, n-Si(111) $700\rm\;\Omega cm$ wafers were flash annealed and an equivalent amount of $10\rm\;nm$ of Ag was deposited at $470^{\circ}\rm C$ following the procedure outlined in ref. \onlinecite{Bannani2008}. The surface of the prepared sample is known to have a high 2D conductivity while the bulk of the sample is almost insulating compared to the surface \cite{Druga2010,Bannani2008}.

As a first test, the performance of the potentiometry feedback is demonstrated by measurement of thermoelectric effects. The thermovoltage between the scanning tip and the sample surface is usually undesired in transport measurements and the identification and reduction of these effects gives a good first insight in the investigated sample system and hinder unwanted effects in later transport measurements. 

To measure the thermovoltage, a potentiometry scan is performed during which the sample is on ground potential and no transport field i.e. lateral current is applied. The result of such a scan is shown in FIG.~\ref{5} (a) and (b) as topography and potential map, respectively. {In the topography, we observe a one atomic layer high island and domain boundaries between different domains of the \AgSi\ reconstruction on the otherwise flat surface}. The corresponding potential map shows an increased potentiometry signal located at the step edges of the island and the domain boundaries, in agreement with the literature \cite{Druga2010}.

The minimum resolution of the potentiometry is determined as the standard deviation of the measured potential of a featureless surface area (square in FIG.~\ref{5} (b) and respective histogram in FIG.~\ref{5} (e)), in this case $\sigma=4.7\rm\;$\textmu V. Note that for this measurement a low bandwidth of $\Delta f=2\rm\;Hz$ was used resulting in a fundamental noise limit to $V_{\rm Limit}(2\rm\;Hz)=1.9\;$\textmu V. In a next step, to verify the observation of the thermovoltage features the tip was heated by a laser ($\lambda=532\rm\;nm,P=20\rm\;mW$) and the measurement was repeated. A strong increase in the potentiometry signal by a factor of around 10 is observed, confirming its nature to be due to thermoelectric effects as seen in the topography and potential map in FIG.~\ref{5} (c) and (d), respectively \cite{footnote3}.
{In the potential map (FIG.~\ref{5} (d)) oscillations in the sample potential can be observed. These oscillations appear pronounced in FIG.~\ref{5} (f) and can be identified as a standing wave pattern which can be explained by thermoelectric effects \cite{Druga2010}}.

When two materials are in tunneling contact, the thermovoltage present at the tunneling junction is proportional to the difference of the squared temperatures of the two materials \cite{Stoevneng1990}. The leading term of the local thermovoltage is given by
\[
V_{\rm th}(x,y)\propto(T_{\rm t}^2-T_{\rm s}^2)\left.\left(\frac{1}{\rho_{\rm s} (x,y,E)}\cdot\frac{\partial\rho_{\rm s} (x,y,E)}{\partial E}\right)\right|_{E=E_{\rm F}}
\]
where $T_{\rm t}$ and $T_{\rm s}$ are the temperature of the tip and sample, respectively. The observed standing wave pattern result from modulations of the local density of states in the sample $\rho_{\rm s}(x,y)$ and were below the resolution of the setup without laser heating of the tip \cite{Druga2010}. The resolution of a standing wave pattern confirms the proper function of the potentiometry feedback.

\section{Determination of step and terrace resistivity}
\subsection{\AgSi}
 \begin{figure*}[!ht]
  \includegraphics{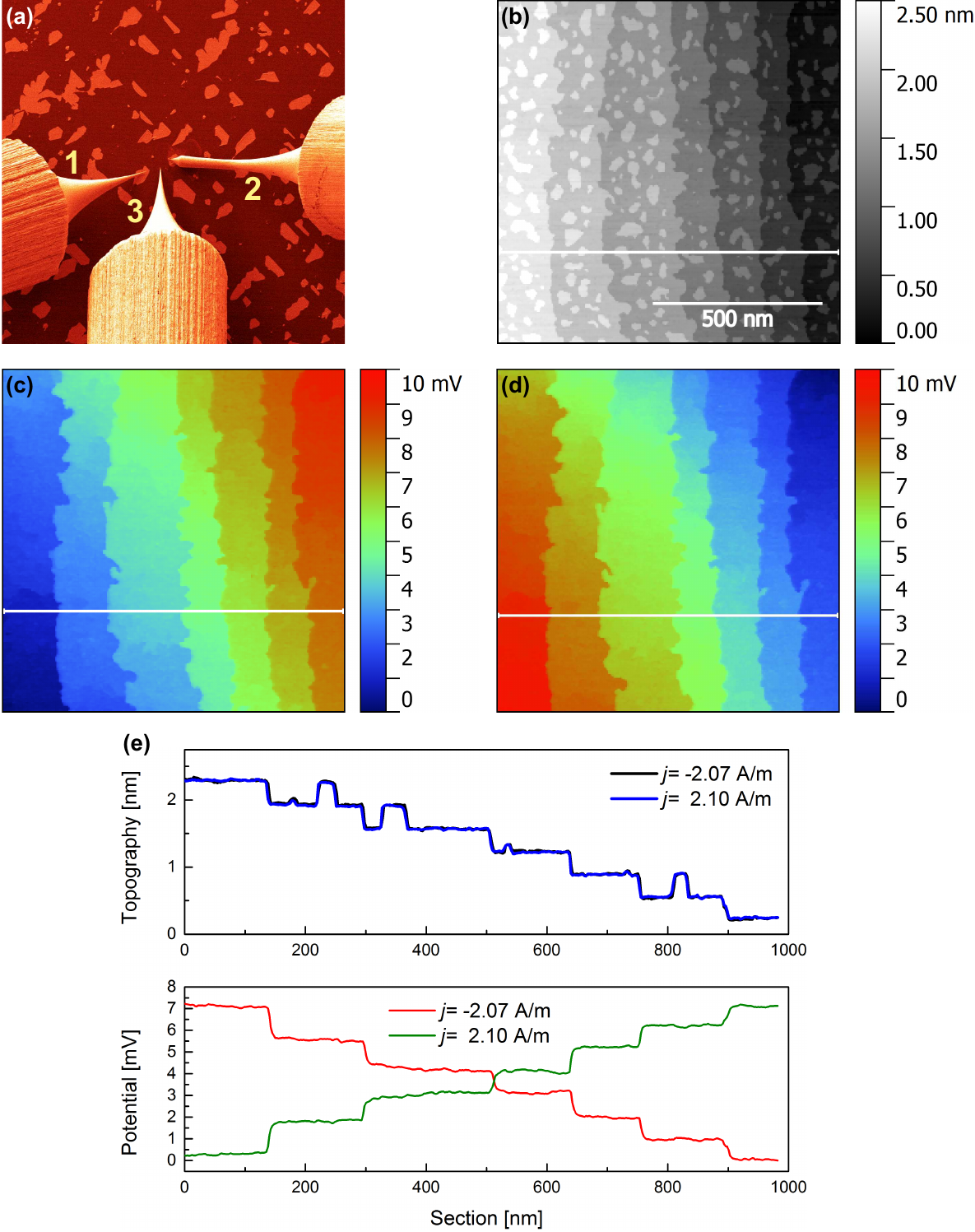}  
  \caption{(Color online) Transport measurements on \AgSi. (a) SEM image of the measurement setup with tip 1 and 2 contacting two 3D Ag islands on the sample surface and tip 3 scanning in between. The distance between the two contacted islands is $d=30(1)\rm\;\text{\textmu} m$. (b) Topography of the \AgSi\ surface. Parallel steps with islands on the terraces can be observed. (c) and (d) Corresponding potential maps for two opposite current directions. The current densities are $j=2.10(8)\rm\;A/m$ and $j=-2.07(8)\rm\;A/m$, respectively. Large voltage drops at the step edges can be observed while there is only a small voltage gradient across the terraces. The one atomic layer high islands show no direct effect on the surrounding potential landscape. (e) The line profiles indicated in (b)-(d) are each averaged over 3 lines. The average voltage drop at the steps is determined to be $\Delta V_{\rm step}=1.08(3)\rm\;mV$. The average slope on the terraces is $E=0.67(8) \rm mV/{\text{\textmu} m}$. Tunneling conditions $V_{\rm offset}=500\rm\;mV$ and $I_{\rm t}=0.5\rm\;nA$. The STP scans each took $4.5\rm\;h$ to perform.}
  \label{6}
  \end{figure*}
To perform transport measurements a lateral current is applied through the sample surface. {A peculiarity of the \AgSi\ surface prepared as described above are the 3D bulk Ag islands visible in the SEM images (FIG.~\ref{6} (a))\cite{Bannani2008}. The bulk Ag island thickness is much larger than that of the actual \AgSi\ reconstruction and can be easily contacted allowing a stable injection of a lateral current.} As a result, for this measurement the optional voltage reference probe is not used. To contact the islands the current injecting tips are approached into tunneling contact and then advanced by few tens of nm with the tunneling feedback deactivated pressing the tips into the islands. Subsequently, the scanning tip is approached to the sample surface as described above and a potentiometry scan is performed. The results of the STP measurements for two opposing directions of the injected current are shown in FIG.~\ref{6} (b-e).

The topography of the \AgSi\ surface (FIG.~\ref{6} (b)) is dominated by approximately parallel monoatomic substrate steps ($h\approx3.1\;$\AA) with an average terrace width of $170\rm\;nm$. One atomic layer high islands are observed on the terraces. The cross sections in FIG.~\ref{6} (e) shows large voltage drops at positions of steps in the topography. Across the terraces an almost linear voltage gradient can be observed with no direct influence of the ad-islands on the voltage drop. {A detailed analysis shows that domain boundaries between different domains of the $(\sqrt3\times\sqrt3)\rm R30^{\circ}$\ reconstruction provide the largest contribution to the voltage drop across the terraces besides the gradient resulting from the undisturbed surface. This observation is in agreement with literature \cite{Bannani2008}}. For reverse current directions we observe an inverted voltage drop indicating a symmetric current response.

The given sample topography of a parallel step array allows a direct evaluation of the step resistivity and terrace resistivity which was not the case in previous STP studies of the \AgSi\ system \cite{Bannani2008, Druga2010}.

From the potential cross sections (averaged over 3 lines) the average voltage drop located at the topographic steps is determined to be $\Delta V_{\rm step}=1.08(3)\rm\;mV$. The current density at the position of the scanning tip is $j=\frac{2I}{\pi d}=2.10(8)\rm\;A/m$ which is determined from the measurement geometry where $d=30(1)\rm\;$\textmu m is the distance of the injection tips and $I$ the absolute current injected into the sample \cite{Ji2011}. From this, the step resistivity can be determined to be $\rho_{\rm step}=\Delta V_{\rm step}/j=514(23)\rm\;$\textmu$\rm\Omega m$. The voltage drop across the terraces is determined from the residual resistance after subtracting the sum of the step resistances from the total resistance and gives $E=0.67(8)\rm\;mV/{\text{\textmu} m}$ corresponding to a resistance of $\rho_{\rm terrace}=E/j=319(39)\rm\;\Omega/\ensuremath{\Box}$. These values are in agreement with previous values from the literature \cite{Druga2010, Bannani2008}. For the present sample, the steps contribute 91\% of the voltage drop across the \AgSi\ surface while the terraces contribute only to 9\%.

\subsection{\Si}
 \begin{figure}[!h]
 \includegraphics{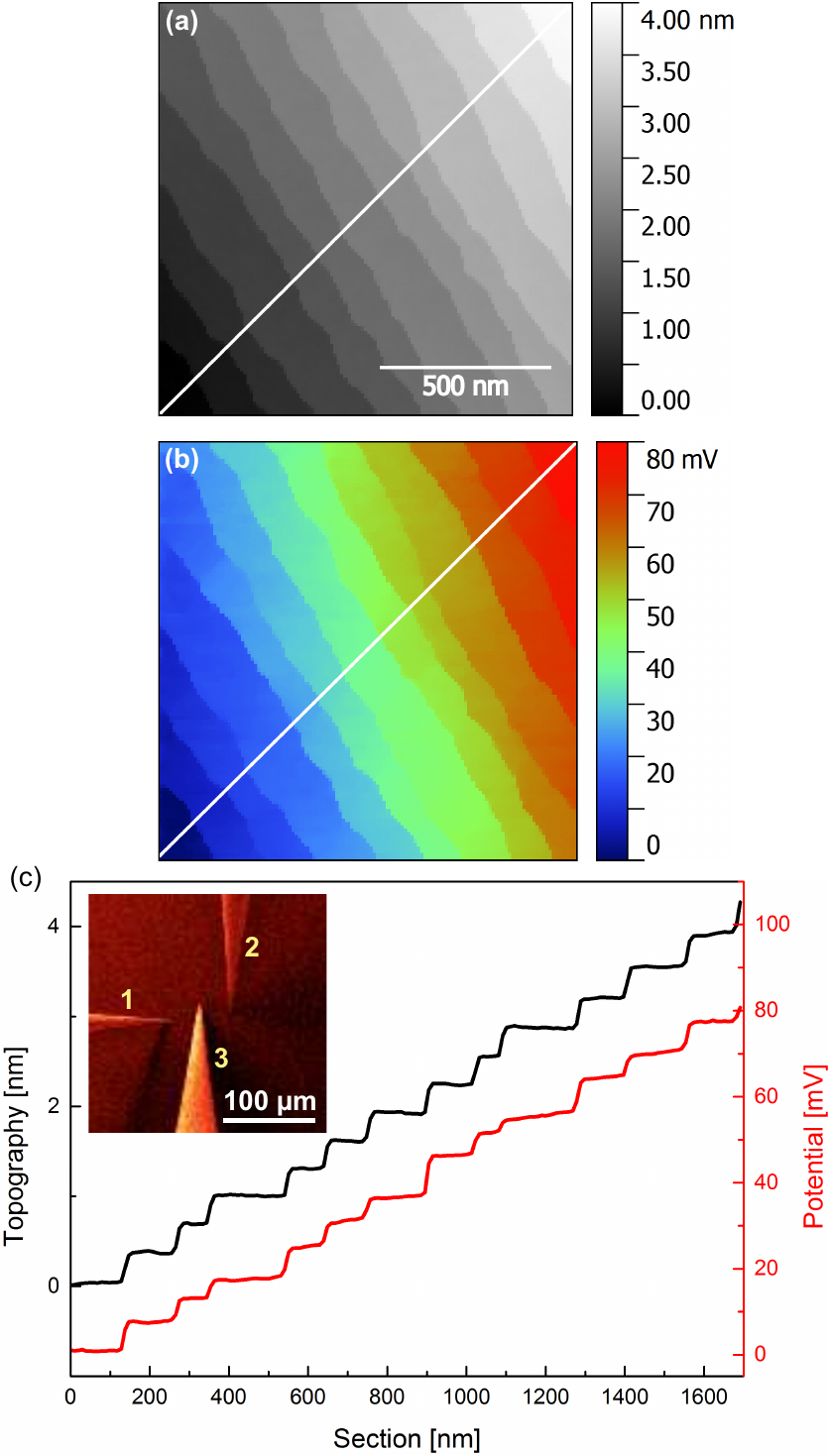}  
 \caption{(Color online) Transport measurements on \Si. (a) Topography of the surface showing parallel steps with an average terraces width of $130\rm\;nm$. (b) Corresponding potential image. A sharp voltage drop is observed at the step edges and a small voltage drop across the terraces. (c) Line profiles indicated in (a) and (b). The average voltage drop at the steps is $\Delta V_{\rm step}=5.67(1.72)\rm\;mV$. The average slope of the terraces is $E=8.33(45)\rm\;mV/{\text{\textmu}m}$. The inset shows the SEM image of the tip setup on the \Si\ surface. The distance between the current injecting tips 1 and 2 is $65\rm\;\mu m$. Tunneling conditions are $V_{\rm offset}=100\rm\;mV$ and $I_{\rm t}=0.01\rm\;nA$.}
 \label{7}
 \end{figure}
 
The \Si\ reconstruction is frequently reported in literature since the beginning of STM studies but is still subject of current investigations \cite{Just2015,Martins2014}. Up to now there are no STP transport studies on this surface reported in literature. For the measurements, we have used the same n-Si(111) $700\rm\;\Omega cm$ substrates as above. The two current injecting tips were brought carefully into direct contact with the freshly prepared \Si\ reconstructed surface (flash annealed to $1230^{\circ}\rm C$) in order to keep the \Si\ terminated surface as intact as possible. As a result, the reference voltage probe is again not necessary.

In contrast to the \AgSi\ the \Si\ reconstruction on n-Si(111) $700\rm\;\Omega cm$ cannot be regarded as a purely 2D conducting material: The bulk and space charge layer of the underlying substrate can have a significant contribution to the measured conductivity \cite{Just2015}. In order to achieve accurate results using STP, a detailed analysis of the conduction through the surface and bulk is mandatory. Such an analysis can be performed by variable distance four-probe measurements as shown recently \cite{Just2015}. A combined analysis of the sample (four-probe measurements + STP) can be performed with the setup presented. {For a current injection tip distance of $d=65(1)\;$\textmu m, at the position of the scanning tip an amount of 69(6)\% of the lateral current is transmitted by the surface conductivity associated with the $(7\times7)$ reconstructed sample surface \cite{Just2015}. The resulting current density at the position of the scanning tip is $j_{\rm surf}=\frac{2I}{\pi d}\cdot0.69(6)=0.056(4)\rm\;A/m$}.

Next, the step and terrace conductivities can be determined as before for the \AgSi\ sample. The corresponding measurements are shown in FIG.~\ref{7}. In the case of \Si\ we observe a variation of the voltage drop at different steps. We attribute this to variations in the shapes and sizes of the terraces and also domain boundaries which results in changes of the current flow paths. The average voltage drop at the steps is determined to be $\Delta V_{\rm step}=5.67(1.72)\rm\;mV$ and across the terraces $E=8.33(45)\rm\;mV/{\text{\textmu} m}$. The resulting resistivity is $\rho_{\rm step}=0.101(32)\rm\;\Omega m$ and $\rho_{\rm terrace}=149(14)\cdot10^3\rm\;\Omega/\ensuremath{\Box}$, respectively while the average step distance in the measurement is $130\rm\;nm$. These results are in agreement with previously reported results \cite{Just2015, Martins2014}. For this surface the steps contribute 82\% of the measured surface resistance while the terraces contribute 18\%.
 
\section{Summary}
We present a multi-tip scanning tunneling potentiometry realization which can be implemented into present multi-tip scanning tunneling microscopes without hardware changes. The resulting setup allows {\it in situ} contacting and STP measurements with flexible positioning of two current injecting electrodes such that the current density in the sample surface and the current direction with respect to the sample under investigation can be controlled with no need of contact patterning.

To map the local potential of the sample on the nanoscale, a third tip is scanning the sample surface between the current injecting electrodes by an interrupted feedback technique, recording the local topography and electric potential. A fourth tip is optional and can be used as a reference voltage probe.

Thermovoltage measurements on the \AgSi\ surface demonstrate the performance of the potentiometry feedback by resolving a standing wave pattern. The resolution of the presented setup is determined to be $\sigma=4.7\rm\;$\textmu V.

The ability to map the local transport field is shown on \AgSi\ where sharp voltage drops located at step edges and symmetrical potential characteristics are observed for reverse current directions. From the investigated sample geometry of a 2D conductor with a parallel step array, conductivities of the terraces and steps can be extracted easily. This straightforward method can further be applied to samples with similar geometry.

For mixed systems of 2D and 3D conduction, a combination of four-probe measurements together with STP allows to first disentangle the bulk and surface conductivity and then determine the contribution of the terraces and steps on the sample surface to the measured conductivity. {This approach is demonstrated and gives a comprehensive view of the conduction in \Si\ on the micro and nanoscale and can be further applied to a wide range of samples that are not limited to pure 2D conductors on an insulating bulk as the case in previous STP studies}.


\begin{thebibliography}{19}%
\makeatletter
\providecommand \@ifxundefined [1]{%
 \@ifx{#1\undefined}
}%
\providecommand \@ifnum [1]{%
 \ifnum #1\expandafter \@firstoftwo
 \else \expandafter \@secondoftwo
 \fi
}%
\providecommand \@ifx [1]{%
 \ifx #1\expandafter \@firstoftwo
 \else \expandafter \@secondoftwo
 \fi
}%
\providecommand \natexlab [1]{#1}%
\providecommand \enquote  [1]{``#1''}%
\providecommand \bibnamefont  [1]{#1}%
\providecommand \bibfnamefont [1]{#1}%
\providecommand \citenamefont [1]{#1}%
\providecommand \href@noop [0]{\@secondoftwo}%
\providecommand \href [0]{\begingroup \@sanitize@url \@href}%
\providecommand \@href[1]{\@@startlink{#1}\@@href}%
\providecommand \@@href[1]{\endgroup#1\@@endlink}%
\providecommand \@sanitize@url [0]{\catcode `\\12\catcode `\$12\catcode
  `\&12\catcode `\#12\catcode `\^12\catcode `\_12\catcode `\%12\relax}%
\providecommand \@@startlink[1]{}%
\providecommand \@@endlink[0]{}%
\providecommand \url  [0]{\begingroup\@sanitize@url \@url }%
\providecommand \@url [1]{\endgroup\@href {#1}{\urlprefix }}%
\providecommand \urlprefix  [0]{URL }%
\providecommand \Eprint [0]{\href }%
\providecommand \doibase [0]{http://dx.doi.org/}%
\providecommand \selectlanguage [0]{\@gobble}%
\providecommand \bibinfo  [0]{\@secondoftwo}%
\providecommand \bibfield  [0]{\@secondoftwo}%
\providecommand \translation [1]{[#1]}%
\providecommand \BibitemOpen [0]{}%
\providecommand \bibitemStop [0]{}%
\providecommand \bibitemNoStop [0]{.\EOS\space}%
\providecommand \EOS [0]{\spacefactor3000\relax}%
\providecommand \BibitemShut  [1]{\csname bibitem#1\endcsname}%
\let\auto@bib@innerbib\@empty
\bibitem [{\citenamefont {Hobara}\ \emph {et~al.}(2007)\citenamefont {Hobara},http://dx.doi.org/10.1063/1.2735593
  \citenamefont {Nagamura}, \citenamefont {Hasegawa}, \citenamefont {Matsuda},
  \citenamefont {Yamamoto}, \citenamefont {Miyatake},\ and\ \citenamefont
  {Nagamura}}]{Hobara2007}%
  \BibitemOpen
  \bibfield  {author} {\bibinfo {author} {\bibfnamefont {R.}~\bibnamefont
  {Hobara}}, \bibinfo {author} {\bibfnamefont {N.}~\bibnamefont {Nagamura}},
  \bibinfo {author} {\bibfnamefont {S.}~\bibnamefont {Hasegawa}}, \bibinfo
  {author} {\bibfnamefont {I.}~\bibnamefont {Matsuda}}, \bibinfo {author}
  {\bibfnamefont {Y.}~\bibnamefont {Yamamoto}}, \bibinfo {author}
  {\bibfnamefont {Y.}~\bibnamefont {Miyatake}}, \ and\ \bibinfo {author}
  {\bibfnamefont {T.}~\bibnamefont {Nagamura}},\ }\href {\doibase 10.1063/1.2735593} {\bibfield
  {journal} {\bibinfo  {journal} {Review of Scientific Instruments}\ }\textbf
  {\bibinfo {volume} {78}},\ \bibinfo {pages} {053705} (\bibinfo {year} {2007})}\BibitemShut {NoStop}%
\bibitem [{\citenamefont {Kim}\ \emph {et~al.}(2007)\citenamefont {Kim},
  \citenamefont {Wang}, \citenamefont {Wendelken}, \citenamefont {Weitering},
  \citenamefont {Li},\ and\ \citenamefont {Li}}]{Kim2007}%
  \BibitemOpen
  \bibfield  {author} {\bibinfo {author} {\bibfnamefont {T.-H.}\ \bibnamefont
  {Kim}}, \bibinfo {author} {\bibfnamefont {Z.}~\bibnamefont {Wang}}, \bibinfo
  {author} {\bibfnamefont {J.~F.}\ \bibnamefont {Wendelken}}, \bibinfo {author}
  {\bibfnamefont {H.~H.}\ \bibnamefont {Weitering}}, \bibinfo {author}
  {\bibfnamefont {W.}~\bibnamefont {Li}}, \ and\ \bibinfo {author}
  {\bibfnamefont {A.-P.}\ \bibnamefont {Li}},\ }\href {\doibase 10.1063/1.2821610} {\bibfield
  {journal} {\bibinfo  {journal} {Review of Scientific Instruments}\ }\textbf
  {\bibinfo {volume} {78}},\ \bibinfo {pages} {123701} (\bibinfo {year} {2007})}\BibitemShut {NoStop}%
\bibitem [{\citenamefont {Nakayama}\ \emph {et~al.}(2012)\citenamefont
  {Nakayama}, \citenamefont {Kubo}, \citenamefont {Shingaya}, \citenamefont
  {Higuchi}, \citenamefont {Hasegawa}, \citenamefont {Jiang}, \citenamefont
  {Okuda}, \citenamefont {Kuwahara}, \citenamefont {Takami},\ and\
  \citenamefont {Aono}}]{Nakayama2012}%
  \BibitemOpen
  \bibfield  {author} {\bibinfo {author} {\bibfnamefont {T.}~\bibnamefont
  {Nakayama}}, \bibinfo {author} {\bibfnamefont {O.}~\bibnamefont {Kubo}},
  \bibinfo {author} {\bibfnamefont {Y.}~\bibnamefont {Shingaya}}, \bibinfo
  {author} {\bibfnamefont {S.}~\bibnamefont {Higuchi}}, \bibinfo {author}
  {\bibfnamefont {T.}~\bibnamefont {Hasegawa}}, \bibinfo {author}
  {\bibfnamefont {C.-S.}\ \bibnamefont {Jiang}}, \bibinfo {author}
  {\bibfnamefont {T.}~\bibnamefont {Okuda}}, \bibinfo {author} {\bibfnamefont
  {Y.}~\bibnamefont {Kuwahara}}, \bibinfo {author} {\bibfnamefont
  {K.}~\bibnamefont {Takami}}, \ and\ \bibinfo {author} {\bibfnamefont
  {M.}~\bibnamefont {Aono}},\ }\href {\doibase 10.1002/adma.201200257}
  {\bibfield  {journal} {\bibinfo  {journal} {Advanced Materials}\ }\textbf
  {\bibinfo {volume} {24}},\ \bibinfo {pages} {1675} (\bibinfo {year}
  {2012})}\BibitemShut {NoStop}%
\bibitem [{\citenamefont {Bannani}, \citenamefont {Bobisch},\ and\
  \citenamefont {M\"oller}(2008)}]{Bannani2008}%
  \BibitemOpen
  \bibfield  {author} {\bibinfo {author} {\bibfnamefont {A.}~\bibnamefont
  {Bannani}}, \bibinfo {author} {\bibfnamefont {C.~A.}\ \bibnamefont
  {Bobisch}}, \ and\ \bibinfo {author} {\bibfnamefont {R.}~\bibnamefont
  {M\"oller}},\ }\href {\doibase 10.1063/1.2968111} {\bibfield  {journal} {\bibinfo  {journal} {Review
  of Scientific Instruments}\ }\textbf {\bibinfo {volume} {79}},\ \bibinfo {pages} {083704} (\bibinfo
  {year} {2008})}\BibitemShut {NoStop}%
\bibitem [{\citenamefont {Pelz}\ and\ \citenamefont {Koch}(1989)}]{Pelz1989}%
  \BibitemOpen
  \bibfield  {author} {\bibinfo {author} {\bibfnamefont {J.~P.}\ \bibnamefont
  {Pelz}}\ and\ \bibinfo {author} {\bibfnamefont {R.~H.}\ \bibnamefont
  {Koch}},\ }\href {\doibase 10.1063/1.1140428}{\bibfield  {journal} {\bibinfo  {journal} {Review
  of Scientific Instruments}\ }\textbf {\bibinfo {volume} {79}},\ \bibinfo {pages} {301} (\bibinfo {year} {1989})}\BibitemShut
  {NoStop}%
\bibitem [{\citenamefont {Druga}\ \emph {et~al.}(2010)\citenamefont {Druga},
  \citenamefont {Wenderoth}, \citenamefont {Homoth}, \citenamefont
  {Schneider},\ and\ \citenamefont {Ulbrich}}]{Druga2010}%
  \BibitemOpen
  \bibfield  {author} {\bibinfo {author} {\bibfnamefont {T.}~\bibnamefont
  {Druga}}, \bibinfo {author} {\bibfnamefont {M.}~\bibnamefont {Wenderoth}},
  \bibinfo {author} {\bibfnamefont {J.}~\bibnamefont {Homoth}}, \bibinfo
  {author} {\bibfnamefont {M.~A.}\ \bibnamefont {Schneider}}, \ and\ \bibinfo
  {author} {\bibfnamefont {R.~G.}\ \bibnamefont {Ulbrich}},\ }\href {\doibase 10.1063/1.3469809}
  {\bibfield  {journal} {\bibinfo  {journal} {Review of Scientific
  Instruments}\ }\textbf {\bibinfo {volume} {81}},\ \bibinfo {pages} {083704} (\bibinfo {year}
  {2010})}\BibitemShut {NoStop}%
\bibitem [{\citenamefont {Feenstra}\ and\ \citenamefont
  {Briner}(1998)}]{Feenstra1998}%
  \BibitemOpen
  \bibfield  {author} {\bibinfo {author} {\bibfnamefont {R.~M.}\ \bibnamefont
  {Feenstra}}\ and\ \bibinfo {author} {\bibfnamefont {B.~G.}\ \bibnamefont
  {Briner}},\ }\href {\doibase 10.1006/spmi.1997.0533} {\bibfield  {journal} {\bibinfo  {journal}
  {Superlattices and Microstructures}\ }\textbf {\bibinfo {volume} {23}},\ \bibinfo {pages} {699}  (\bibinfo {year} {1998})}\BibitemShut {NoStop}%
\bibitem [{\citenamefont {Willke}\ \emph {et~al.}(2015)\citenamefont {Willke},
  \citenamefont {Druga}, \citenamefont {Ulbrich}, \citenamefont {Schneider},\
  and\ \citenamefont {Wenderoth}}]{Willke2015}%
  \BibitemOpen
  \bibfield  {author} {\bibinfo {author} {\bibfnamefont {P.}~\bibnamefont
  {Willke}}, \bibinfo {author} {\bibfnamefont {T.}~\bibnamefont {Druga}},
  \bibinfo {author} {\bibfnamefont {R.~G.}\ \bibnamefont {Ulbrich}}, \bibinfo
  {author} {\bibfnamefont {M.~A.}\ \bibnamefont {Schneider}}, \ and\ \bibinfo
  {author} {\bibfnamefont {M.}~\bibnamefont {Wenderoth}},\ }\href {\doibase 10.1038/ncomms7399}
  {\bibfield  {journal} {\bibinfo  {journal} {Nature Communications}\ }\textbf
  {\bibinfo {volume} {6}},\ \bibinfo {pages} {6399} (\bibinfo {year} {2015})}\BibitemShut {NoStop}%
\bibitem [{\citenamefont {Muralt}\ and\ \citenamefont
  {Pohl}(1986)}]{Muralt1986}%
  \BibitemOpen
  \bibfield  {author} {\bibinfo {author} {\bibfnamefont {P.}~\bibnamefont
  {Muralt}}\ and\ \bibinfo {author} {\bibfnamefont {D.~W.}\ \bibnamefont
  {Pohl}},\ }\href {\doibase 10.1063/1.96491} {\bibfield  {journal} {\bibinfo  {journal} {Applied
  Physics Letters}\ }\textbf {\bibinfo {volume} {48}},\ \bibinfo {pages} {514} (\bibinfo {year}
  {1986})}\BibitemShut {NoStop}%
\bibitem [{\citenamefont {Rozler}\ and\ \citenamefont
  {Beasley}(2008)}]{Rozler2008}%
  \BibitemOpen
  \bibfield  {author} {\bibinfo {author} {\bibfnamefont {M.}~\bibnamefont
  {Rozler}}\ and\ \bibinfo {author} {\bibfnamefont {M.~R.}\ \bibnamefont
  {Beasley}},\ }\href {\doibase 10.1063/1.2953097} {\bibfield  {journal} {\bibinfo  {journal}
  {Review of Scientific Instruments }\ }\textbf {\bibinfo {volume} {79}},\ \bibinfo {pages} {073904}  (\bibinfo {year} {2008})}\BibitemShut {NoStop}%
\bibitem [{\citenamefont {Wang}\ \emph {et~al.}(2013)\citenamefont {Wang},
  \citenamefont {Munakata}, \citenamefont {Rozler},\ and\ \citenamefont
  {Beasley}}]{Wang2013}%
  \BibitemOpen
  \bibfield  {author} {\bibinfo {author} {\bibfnamefont {W.}~\bibnamefont
  {Wang}}, \bibinfo {author} {\bibfnamefont {K.}~\bibnamefont {Munakata}},
  \bibinfo {author} {\bibfnamefont {M.}~\bibnamefont {Rozler}}, \ and\ \bibinfo
  {author} {\bibfnamefont {M.~R.}\ \bibnamefont {Beasley}},\ }\href {\doibase 10.1103/PhysRevLett.110.236802}
  {\bibfield  {journal} {\bibinfo  {journal} {Physical Review Letters}\
  }\textbf {\bibinfo {volume} {110}},\ \bibinfo {pages} {236802} (\bibinfo {year} {2013})}\BibitemShut
  {NoStop}%
\bibitem [{\citenamefont {Just}\ \emph {et~al.}(2015)\citenamefont {Just},
  \citenamefont {Blab}, \citenamefont {Korte}, \citenamefont {Cherepanov},
  \citenamefont {Soltner},\ and\ \citenamefont {Voigtl\"ander}}]{Just2015}%
  \BibitemOpen
  \bibfield  {author} {\bibinfo {author} {\bibfnamefont {S.}~\bibnamefont
  {Just}}, \bibinfo {author} {\bibfnamefont {M.}~\bibnamefont {Blab}}, \bibinfo
  {author} {\bibfnamefont {S.}~\bibnamefont {Korte}}, \bibinfo {author}
  {\bibfnamefont {V.}~\bibnamefont {Cherepanov}}, \bibinfo {author}
  {\bibfnamefont {H.}~\bibnamefont {Soltner}}, \ and\ \bibinfo {author}
  {\bibfnamefont {B.}~\bibnamefont {Voigtl\"ander}},\ }\href {\doibase 10.1103/PhysRevLett.115.066801} {\bibfield
  {journal} {\bibinfo  {journal} {Physical Review Letters}\ }\textbf {\bibinfo
  {volume} {115}},\ \bibinfo {pages} {066801} (\bibinfo {year} {2015})}\BibitemShut {NoStop}%
\bibitem [{\citenamefont {Kanagawa}\ \emph {et~al.}(2003)\citenamefont
  {Kanagawa}, \citenamefont {Hobara}, \citenamefont {Matsuda}, \citenamefont
  {Tanikawa}, \citenamefont {Natori},\ and\ \citenamefont
  {Hasegawa}}]{Kanagawa2003}%
  \BibitemOpen
  \bibfield  {author} {\bibinfo {author} {\bibfnamefont {T.}~\bibnamefont
  {Kanagawa}}, \bibinfo {author} {\bibfnamefont {R.}~\bibnamefont {Hobara}},
  \bibinfo {author} {\bibfnamefont {I.}~\bibnamefont {Matsuda}}, \bibinfo
  {author} {\bibfnamefont {T.}~\bibnamefont {Tanikawa}}, \bibinfo {author}
  {\bibfnamefont {A.}~\bibnamefont {Natori}}, \ and\ \bibinfo {author}
  {\bibfnamefont {S.}~\bibnamefont {Hasegawa}},\ }\href {\doibase 10.1103/PhysRevLett.91.036805} {\bibfield
  {journal} {\bibinfo  {journal} {Physical Review Letters}\ }\textbf {\bibinfo
  {volume} {91}},\ \bibinfo {pages} {036805} (\bibinfo {year} {2003})}\BibitemShut {NoStop}%
\bibitem [{\citenamefont {Cherepanov}\ \emph {et~al.}(2012)\citenamefont
  {Cherepanov}, \citenamefont {Zubkov}, \citenamefont {Junker}, \citenamefont
  {Korte}, \citenamefont {Blab}, \citenamefont {Coenen},\ and\ \citenamefont
  {Voigtl\"ander}}]{Cherepanov2012}%
  \BibitemOpen
  \bibfield  {author} {\bibinfo {author} {\bibfnamefont {V.}~\bibnamefont
  {Cherepanov}}, \bibinfo {author} {\bibfnamefont {E.}~\bibnamefont {Zubkov}},
  \bibinfo {author} {\bibfnamefont {H.}~\bibnamefont {Junker}}, \bibinfo
  {author} {\bibfnamefont {S.}~\bibnamefont {Korte}}, \bibinfo {author}
  {\bibfnamefont {M.}~\bibnamefont {Blab}}, \bibinfo {author} {\bibfnamefont
  {P.}~\bibnamefont {Coenen}}, \ and\ \bibinfo {author} {\bibfnamefont
  {B.}~\bibnamefont {Voigtl\"ander}},\ }\href {\doibase 10.1063/1.3694990} {\bibfield  {journal}
  {\bibinfo  {journal} {Review of Scientific Instruments}\ }\textbf {\bibinfo
  {volume} {83}},\ \bibinfo {pages} {033707} (\bibinfo {year} {2012})}\BibitemShut {NoStop}%
\bibitem [{\citenamefont {mProbes GmbH}(2015)}]{mProbes2015}%
  \BibitemOpen
  \bibfield  {author} {\bibinfo {author} {\bibnamefont {mProbes GmbH}},\ }\href
  {http://www.mprobes.com/} {\enquote {\bibinfo {title} {http://www.mprobes.com/}}\ } (\bibinfo {year}
  {2015})\BibitemShut {NoStop}%
\bibitem [{\citenamefont {FEMTO}(2015)}]{Femto2015}%
  \BibitemOpen
  \bibfield  {author} {\bibinfo {author} {\bibnamefont {FEMTO Messtechnik
    GmbH}},\ }\href
  {http://www.femto.de/en/} {\enquote {\bibinfo {title} {http://www.femto.de/en/}}\ } (\bibinfo {year} {2015})\BibitemShut {NoStop}%
\bibitem{footnote1}Due to the voltage spikes occurring when the potentiometry bias voltage is set before the potentiometry feedback and when the topography bias voltage is set before re-activating the topography feedback a short delay time of a few ms is recommended to be implemented respectively to allow the system to settle.   
\bibitem{footnote2}If the voltage probe is not present, it is assumed that the transport field is symmetrical, with the option to correct the voltages later. 
\bibitem [{\citenamefont {Hamada}\ and\ \citenamefont
  {Hasegawa}(2012)}]{Hamada2012}%
  \BibitemOpen
  \bibfield  {author} {\bibinfo {author} {\bibfnamefont {M.}~\bibnamefont
  {Hamada}}\ and\ \bibinfo {author} {\bibfnamefont {Y.}~\bibnamefont
  {Hasegawa}},\ }\href {\doibase 10.1143/JJAP.51.125202} {\bibfield  {journal} {\bibinfo  {journal}
  {Japanese Journal of Applied Physics}\ }\textbf {\bibinfo {volume} {51}},\ \bibinfo {pages} {125202}
  (\bibinfo {year} {2012})}\BibitemShut {NoStop}%
\bibitem{footnote3}Surface photoelectric effects were ruled out by switching off the laser for a short time and still observing the increased potentiometry signal.
\bibitem [{\citenamefont {St{\o}vneng}\ and\ \citenamefont
  {Lipavsk{\'y}}(1990)}]{Stoevneng1990}%
  \BibitemOpen
  \bibfield  {author} {\bibinfo {author} {\bibfnamefont {J.~A.}\ \bibnamefont
  {St{\o}vneng}}\ and\ \bibinfo {author} {\bibfnamefont {P.}~\bibnamefont
  {Lipavsk{\'y}}},\ }\href {\doibase 10.1103/PhysRevB.42.9214} {\bibfield  {journal} {\bibinfo  {journal}
  {Physical Review B}\ }\textbf {\bibinfo {volume} {42}},\ \bibinfo {pages} {9214} (\bibinfo {year}
  {1990})}\BibitemShut {NoStop}%
\bibitem [{\citenamefont {Ji}\ \emph {et~al.}(2011)\citenamefont {Ji},
  \citenamefont {Hannon}, \citenamefont {Tromp}, \citenamefont {Perebeinos},
  \citenamefont {Tersoff},\ and\ \citenamefont {Ross}}]{Ji2011}%
  \BibitemOpen
  \bibfield  {author} {\bibinfo {author} {\bibfnamefont {S.-H.}\ \bibnamefont
  {Ji}}, \bibinfo {author} {\bibfnamefont {J.~B.}\ \bibnamefont {Hannon}},
  \bibinfo {author} {\bibfnamefont {R.~M.}\ \bibnamefont {Tromp}}, \bibinfo
  {author} {\bibfnamefont {V.}~\bibnamefont {Perebeinos}}, \bibinfo {author}
  {\bibfnamefont {J.}~\bibnamefont {Tersoff}}, \ and\ \bibinfo {author}
  {\bibfnamefont {F.~M.}\ \bibnamefont {Ross}},\ }\href {\doibase 10.1038/NMAT3170} {\bibfield
  {journal} {\bibinfo  {journal} {Nature Materials}\ }\textbf {\bibinfo
  {volume} {11}},\ \bibinfo {pages} {114} (\bibinfo {year} {2011})}\BibitemShut {NoStop}%
\bibitem [{\citenamefont {Martins}\ \emph {et~al.}(2014)\citenamefont {Martins},
    \citenamefont {Smeu}, \citenamefont {Livadaru}, \citenamefont {Guo},\ and\ \citenamefont {Wolkow}}]{Martins2014}%
    \BibitemOpen
    \bibfield  {author} {\bibinfo {author} {\bibfnamefont {B. V. C.}~\bibnamefont
    {Martins}}, \bibinfo {author} {\bibfnamefont {M.}~\bibnamefont {Smeu}}, \bibinfo
    {author} {\bibfnamefont {L.}~\bibnamefont {Livadaru}}, \bibinfo {author}
    {\bibfnamefont {H.}~\bibnamefont {Guo}},\ and\ \bibinfo {author}
    {\bibfnamefont {R. A.}~\bibnamefont {Wolkow}},\ }\href {\doibase 10.1103/PhysRevLett.112.246802} {\bibfield
    {journal} {\bibinfo  {journal} {Physical Review Letters}\ }\textbf {\bibinfo
    {volume} {112}},\ \bibinfo {pages} {246802} (\bibinfo {year} {2014})}\BibitemShut {NoStop}%
  \end{thebibliography}

\end{document}